\documentclass[11pt]{article}
\usepackage{Blois,epsfig}
\usepackage{amssymb}

\def\pom{I\!\!P}
\def\dpe{DI\!\!PE}

\def\be{\begin{equation}}
\def\ee{\end{equation}}
\def\bea{\begin{eqnarray}}
\def\eea{\end{eqnarray}}

\begin{document}
\begin{flushright}
FERMILAB-CONF-05-412-E
\end{flushright}
\vspace*{1cm}
\begin{center}
\Large{\textbf{XIth International Conference on\\ Elastic and Diffractive
Scattering\\ Ch\^{a}teau de Blois, France, May 15 - 20, 2005}}
\end{center}

\vspace*{2cm}
\title{QCD AND HARD DIFFRACTION AT THE LHC}

\author{ M.G. ALBROW }

\address{Fermi National Accelerator Laboratory\\
Wilson Road, Batavia, IL 60510, USA}

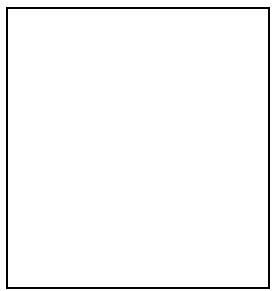
\maketitle\abstracts{
As an introduction to QCD at the LHC I give an overview of QCD at the Tevatron,
emphasizing the high $Q^2$ frontier which will be taken over by the LHC. After
describing briefly the LHC detectors I discuss high mass diffraction, in
particular central exclusive production of Higgs and vector boson pairs. I
introduce the FP420 project to measure the scattered protons 420m
downstream of ATLAS and CMS.}

\section{Recent QCD Studies at the Tevatron (CDF and DO$\!\!\!\!/$)}

\subsection{Jets}

Every $pp$ and $p\bar{p}$ interaction is QCD! (Even two photon processes
$p+p \rightarrow p\oplus e^+e^- \oplus p$ and $p+p \rightarrow p\oplus W^+W^- \oplus p$,
where $\oplus$ means a large rapidity gap with \emph{no particles} have small QCD corrections!) 
From May 2003 to May 2005 CDF (DO$\!\!\!\!/$) published 43
(10) ``QCD" papers on many topics \cite{cdf,dzero}. All of these will be studied at the LHC, most in the early
low-luminosity period.

    Most obviously QCD is the high $p_T$ jet (J) production cross section, its $E_T,\eta$ and
$\sqrt{s}$ dependence. The Tevatron took data at $\sqrt{s}$ = 630, 1800 and 1960 
GeV. Hopefully the LHC will take some data below 14 TeV ... 1960 GeV would be useful to
compare $pp$ and $p\bar{p}$, albeit with different detectors. Hadron jets are not uniquely 
defined objects; we need some
\emph{algorithm}. Good algorithms give a good approximation to a scattered parton's
4-momentum (itself not well-defined quantum mechanically) and allow reasonable
comparisons between experimental data and a theoretically-based simulation. Most
commonly used are \emph{cone} (circle in $\eta,\phi$) and $k_T$ (transverse momentum 
w.r.t. a jet axis) algorithms. Internal jet structure (charged multiplicity
distributions, quark/gluon differences, $c$- and $b$-quark content), di-jet azimuthal
separation, and 3-jet events have all been studied, and values of $\alpha(s)$
extracted. At the Tevatron central jets with $E_T$ below 50 GeV are mostly gluon jets;
above 400 GeV (the spectra now extend to $\approx$ 600 GeV
mostly from $q\bar{q}$ scattering).
%
 This $E_T$ dependence enables us to study $q/g$ fragmentation differences.
At the LHC the jets with $E_T \lesssim$ 100 GeV will be nearly pure gluon jets. Events at the
highest $E_T$ are mostly spectacularly clean (on event displays) 2-jet (or sometimes 3-jet)
final states. Projected cross sections at the LHC extend, in 100 fb$^{-1}$, to $E_T \approx$ 
4 TeV ($M_{JJ}$ = 8 TeV)
with still a few events per 100 GeV bin. Let us hope these
projections are quite wrong! Let us hope for peaks or excesses or even a cut-off (due to black hole
production) - the end of high-$p_T$ physics!

     The LHC will extend our coverage in the $(x,Q^2)$ plane ($x$ = Bjorken-$x$) by more than
an order of magnitude to smaller $x$ and higher $Q^2$ ... up to $\approx 10^8$ GeV$^2$
corresponding to a transverse distance $\approx 2\times 10^{-18}$ cm. To reach the smallest
$x$-values, $\lesssim 10^{-5}$, requires very forward detectors like LUCID (ATLAS) and TOTEM
(CMS). Very forward jet measurements allow important studies of BFKL and Mueller-Navelet jets.
A BFKL pomeron is a color-singlet exchange between quarks, constructed from a pair of
reggeized gluons in a ladder. It enhances the cross section for $qq$-scattering especially at
large $\frac{s}{t}$, i.e. for jets with large rapidity separation. It is interesting to study
both the ``elastic" case (with a large rapidity gap $\Delta y$ between them) and the inelastic
case (with $n$ minijets in between). This is a fundamental probe of a new regime:
non-perturbative QCD at short distances; it requires very forward hadron calorimeters in both
forward directions.

\subsection{Particle Production}

Generic particle production has been much studied at the Tevatron, and is important
information for our understanding of non-perturbative QCD. Results on total production
cross sections and/or their $p_T, \eta$ and $\sqrt{s}$ dependence have been published for
strongly interacting particles ($K^o_S ,\Lambda, c(D), J/\psi,\psi', B, B\bar{B}, \Upsilon,
t\bar{t}$) and not-strongly interacting ($\gamma, e^+e^-, W, Z, VV (V = \gamma,W,Z))$. In
all cases the differential production cross sections are compared with Monte Carlo simulations
(which have some QCD-inspired hadronization code) and generally agree within claimed
uncertainties. B-hadrons and $\Upsilon$ are measured from $p_T$ = 0 to 25 (20) GeV/c
respectively, the former thanks to a fast secondary-vertex trigger in CDF.

\subsection{Hadron Decays}

Hadron decays are also of course QCD, albeit non-perturbative, and thanks to the special 
triggers and large
production rate of $B$-hadrons, especially $B_s$ and $B_c$ which are inaccessible at
$B$-factories, new and important physics is done. At the LHC this program
will continue, with LHCB but also with ATLAS and CMS. Results from the Tevatron on
the following decays have been published (lifetimes and branching fractions): $B^\circ_s
\rightarrow \phi\phi, J/\psi \phi; B^\circ \rightarrow hadrons, \gamma X, J/\psi X, J/\psi K^*
\pi^+\pi^-$ and $\tau(\Lambda_b \rightarrow J/\psi \Lambda$). CDF made the first observation
of the decay $B_s \rightarrow \phi\phi$ with 12 events on a background of 2, and
BR($B_s\rightarrow \phi\phi) = [1.4 \pm 0.6 (stat) \pm 0.2 (syst) \pm 0.5 (BRs)] \times
10^{-5}$. The rare decay $B_d \rightarrow \pi^+\pi^-$ has also been observed with a BR $5
\times 10^{-6}$. At least a factor $\times$ 25 in sensitivity is expected from the Tevatron. 

\subsection{Hadron Spectroscopy}

Hadron spectroscopy with $c$- and $b$-quarks provides
important tests especially of Lattice QCD. The hidden charm state $X(3872) \rightarrow J/\psi
\pi^+\pi^-$ was clearly seen in CDF within a week of its discovery by BELLE. More attention
should be paid by Tevatron physicists to its potential in this regard. Pentaquark searches
(e.g. $\Xi^-\pi^-$) proved negative, despite very high sensitivity. A competitive measurement
of $\Delta m (D^+_s - D^+)$ was published. CDF observed $B_c \rightarrow J/\psi \pi^+$ as a
narrow peak at 6.287 $\pm$ 0.005 GeV/c$^2$ (much higher precision than the Run 1 observation
in the semileptonic decay). This state is the most perturbative hadron that does not decay
strongly. Its spectroscopy will be very valuable ($B_c^* \rightarrow B_c\gamma, B_c \pi\pi$).
A possibility is that hadrons with two heavy quarks can be made relatively cleanly in double
pomeron interactions ($\pom (gg) \pom (gg) \rightarrow B_c B D X?$)

\subsection{Probing Very Small-x Gluons}

At very small $x\approx 10^{-5}$, gluon densities become very high
and new saturation phenomena can occur. At HERA $q(x)$ has been measured and
$g(x)$ inferred by evolution and by charm production. At the LHC very low-$x$
gluons can be measured more directly. In a 2-parton scattering process resulting
in $n$ jets with $p_T^i$ and $\eta_i$, the incident parton
$x$'s are given by:
\[x_{1(2)} = \frac{1}{\sqrt{s}}\sum_{i=1}^{n} p_T^i e^{+(-)\eta_i}\]
so e.g. for $\sqrt{s} = 14$ TeV, with two jets with $p_T$ = 5 GeV, $\eta_1 = \eta_2$ = 4 (2.1$^\circ$) we
have $x_1$ = 0.08 and $x_2 = 1.4 \times 10^{-5}$. To reach this physics we need
to instrument the few degree region with tracking and calorimetry (electromagnetic and hadronic)
measuring muons, $J/\psi$, jets and photons. CASTOR (in CMS) and TOTEM do this
only partially.

\subsection{Underlying Event}

In a hard collision producing jets color fields are everywhere and it
is not possible to separate an ``underlying event" from jet fragmentation
products. However we can define a region of solid angle perpendicular
in azimuth to a jet axis and measure charged particle multiplicities and
$p_T$-spectra there, to make comparisons between soft collisions (no high $p_T$
jets), hard collisions and event generators. This is interesting in itself, as
well as affecting the comparisons of jet spectra with perturbative QCD
calculations. A special case is the study of the associated hadrons in events
with one or two $W \rightarrow l\nu$ or $Z\rightarrow l^+l^-$, when the underlying event is unambiguously
defined. CDF has made many studies: e.g. the $p_T$ spectrum at 90$^\circ$ to the
leading jet axis is harder than in minimum bias interactions; the number of
charged particles there is also higher but hardly changes as the leading
(charged) jet increases from $p_T$ = 20 GeV/c to 150 GeV/c. Such observations
allow tuning of Monte Carlo generators (e.g. PYTHIA Tune A) for the Tevatron,
giving us our best estimates of events at LHC.

Part of the underlying event can be double or multiple parton scattering (DPS/MPS). ISR
data hinted at the existence of DPS observing 4 ``jets" that seemed to be more
pair-wise back-to-back than expected from double bremsstrahlung ($2\rightarrow
4$). It has been more clearly seen in UA2 and in CDF (with $\gamma + JJJ$). One
can study specific parton correlations, e.g. if quarks are accompanied
by a correlated gluon cloud we should have a positive correlation between a
Drell-Yan lepton pair and a $c\bar{c}/b\bar{b}$ pair. If di-quarks have some
significance in the proton wave function then double Drell-Yan could be enhanced
in $p\bar{p}$ collisions. 

\subsection{Diffraction and Rapidity Gaps}

    Since May 2003 results from the Tevatron included diffractive production of
di-jets and $J/\psi$, inclusive double pomeron exchange $\dpe$, and double-gap
events $[XGXG\bar{p}]$ where $X$ is ``hadrons" and $G$ is a rapidity gap. The
latter process is like $\dpe$ with one proton dissociating. We showed that if
the price paid for a rapidity gap is about 0.01, the extra price for a second gap is
only about 0.1. Prior to 2003 a wealth of diffractive data has been published
from the Tevatron ($J$ = jet): $\sigma_T, \frac{d\sigma}{dt}, \frac{d\sigma}{dM^2dt}(SDE),
SDE \rightarrow JJ, J/\psi, b, W, Z, \dpe \rightarrow JJ$ and $J-G-J$.

    At the LHC we have a larger rapidity range than at the Tevatron, $\Delta y = 2
ln \frac{\sqrt{s}}{m_p} = 19.2$ (cf. 15.2 at the Tevatron). Noting that as a
rule-of-thumb a hadronic cluster takes up about $\Delta y = 2 ln M(GeV)$ in
rapidity, we can have the following situations: (a) Single Diffractive
Excitation up to $M$ = 2 TeV with a $\Delta y$ = 4 gap (dominated by $\pom$)
(b) Opposite side very forward jets separated by a $\Delta y$
= 6 gap (c) $\dpe$ with a 700 GeV central state and two $\Delta y \gtrsim$ 3
gaps (d) $p-G3-X-G3-X-G3-p$ multi-pomeron exchange with $M(X) \approx 6$ GeV and
3-unit gaps.

\section{At the LHC}

\subsection{Detectors}

Ways in which the Tevatron program and the LHC program can mutually enhance each
other have been studied in the series of TeV4LHC Workshops (google tev4lhc). The
QCD group had several subgroups on pdf's, jet algorithms, matrix element/Monte
Carlos, hadronization and underlying events, and diffraction.

The LHC is a \emph{QCD Machine} and \emph{all} the experiments are \emph{QCD
Detectors}. I will not say more on ALICE, very QCD oriented in its study of
heavy ion collisions and possible QCD phase transitions, and LHCB which will
teach us not only about CP-violation but much about hadron spectroscopy and
decays. The QCD potential of both the major detectors, ATLAS and CMS, will be
greatly enhanced by detectors in the forward and very forward regions. At the
same intersection (P5) as CMS the TOTEM experiment is designed to measure
$\sigma_T$, elastic scattering and single diffraction, with some special
medium/high-$\beta$ LHC operation and hopefully different $\sqrt{s}$ values. Apart from
a series of roman pots out to 215m, the TOTEM T1 and T2 detectors provide some
tracking and calorimetry in the forward regions, and T2 is followed on at least
one side by the CASTOR(CMS) calorimeter. Together TOTEM and CMS attempt to cover
4$\pi$ with detectors and this enables a rich diffractive program. ATLAS does
not have similar coverage; there is to be a forward multi-cell gas Cerenkov counter,
LUCID, and some roman pots, motivated by luminosity measurements but with some diffractive
physics capability. Groups in both CMS (with TOTEM) and ATLAS would like to add very
forward proton detectors, 420m downstream on both sides, a project in the R\&D phase 
called ``FP420" \cite{fp}.

\subsection{Central Exclusive Production}

Central Exclusive Production, $pp\rightarrow p \oplus X \oplus p$, where $X$ is a specific
state, becomes very interesting at the LHC; according to the above rule-of-thumb $M_X(max)
\approx 100$ GeV at the Tevatron becomes $\approx$ 700 GeV at the LHC. We therefore reach
into the domain of Higgs $H$, $W^+W^-$, $ZZ$, $t\bar{t}$ production and maybe the unknown
$X$. The possibility of seeing $pp\rightarrow p \oplus H \oplus p$ is exciting. The main
channel for $H$ production at hadron colliders is $gg$ fusion through a top loop. Another gluon exchange can
cancel the color and even leave the protons in their ground states, to be measured way
downstream (after 116m of 8T dipoles in a vacuum!). The cross section has been estimated \cite{kkmrs} to
be about 3 fb for a SM $H$ of about 130 GeV, giving some 100 events ($\times$ acceptance)
in a 30 fb$^{-1}$ year. There are theoretical uncertainties involving skewed gluon
distributions, gluon $k_T$, gluon radiation (Sudakov form factors) etc, estimated to be
a factor $\approx 2.5$. The theory can be tested at the Tevatron by the same
process with the top loop replaced with a $b(c)$-loop $\rightarrow \chi_{b(c)}$ or a
$u$-loop $\rightarrow \gamma\gamma$. We are looking for these processes and
have good candidates in CDF for exclusive $\chi_c \rightarrow J/\psi \gamma \rightarrow
\mu^+\mu^-\gamma$. Unfortunately we do not have forward proton detection and the
background is not as low as it might have been. If even a dozen or so 
$pp\rightarrow p \oplus H \oplus p$ events are measured it will be important; the mass can
be measured by the missing mass technique \cite{albmm} with $\sigma_M \lesssim$ 2 GeV per event, and it
can be established that the state is a scalar \cite{kmr1} (hard to do another way before the ILC).
Exclusive $\dpe \rightarrow q\bar{q}$ di-jets are strongly suppressed by the $J_Z = 0$
rule \cite{kkmrs}, so the signal:background ratio is high ($\approx 1$). In the case of a non-SM Higgs sector
life can be even more interesting! The production cross section can be much
higher \cite{ell} and one
can have a close triplet $h,A,H$ where $A$ is mostly CP-odd. It can be difficult to
resolve these states if they are within a few GeV. In central exclusive production the
middle state $A$ is absent (CP-odd) and the $h$ and $H$ may be resolved.

   The TOTEM roman pots at 215m have no acceptance, in the standard
low-$\beta$ high luminosity running, for $M_X \lesssim$ 300 GeV. We need to measure
   protons that have lost $\lesssim$ 1\% of their energy. This requires going further
   forward; there is an ideal location 420m downstream. Here there is a 15m straight
   section where a cryogenic by-pass will allow us to put very small (6mm $\times$ 24mm)
   tracking detectors (probably 3D silicon) within 3mm of the proton beam. We plan to also have high
   precision ($\approx$ 10ps) time of flight counters, which measure the interaction point in $z$ from the
   time difference. Measuring
   protons here, on both sides or with one proton in a 220m detector, gives acceptance for
   $M_X \gtrsim$ 50 GeV. A consortium of ATLAS and CMS people submitted a LOI to the LHCC
   in June 2005, to support R\&D towards common technical solutions (FP420)
   \cite{fp}.
   
   \subsection{Vector Boson Pair Production by $\dpe$}
   
   Vector boson pairs are especially interesting whether produced
   non-diffractively or diffractively. Consider prompt pairs, i.e. not from
   $t\bar{t}$ and not from $H$. At the Tevatron 90\% of $W^+W^-$ and $W^\pm Z$
   come from $q\bar{q}$ annihilation with $t$-channel quark exchange. Also
   $q\bar{q}$ annihilation with an $s$-channel $W^*/Z^*$ contributes $\approx$ 10\%
   to $W^+W^-$ and $W^\pm Z$ (not $ZZ$). Any pair (even $W^+W^+$) can be
   produced by two incident quarks radiating virtual $W/Z$ which scatter and
   become real; in this case the quarks give forward high-$p_T$ (``tagging") 
   jets. This is an important process for the quartic boson coupling (and
   direct channel resonances such as the Higgs!). 
   At the Tevatron cross sections (limits for $WZ$ and $ZZ$) agree with the CTEQ
   NLO predictions $\sigma(WW/WZ/ZZ)$ = 12.4/3.65/1.39 pb. At the LHC they are a
   factor $\approx$ 10 higher. If the same rules-of-thumb apply, about 1\% 
   will be single diffractive and 10$^{-3}$ will be in $\dpe$. So our
   ``guesstimate" for $\dpe \rightarrow WW(ZZ)$ + \emph{anything} is $\approx$ 1pb (100fb).
   
   Exclusive $V$-pairs are another matter. The process
   $pp\rightarrow p \oplus W^+W^- \oplus p$ by two photon exchange is guaranteed
   and $\approx$ 100 fb; the protons will have very small (Coulombic) momentum
   transfer $t$ which helps distinguish this from the more interesting two pomeron
   exchange. Exclusive $\dpe \rightarrow W^+W^-$ should be completely 
   negligible, unless Alan White is right~\cite{arw}. The signature is spectacular. In the
   (only) 4.5\% of cases where both $W$ decay to $e$ or $\mu$ the two leptons
   are on a vertex with \emph{no hadrons}. If the two protons are measured there are several
   missing mass variables of interest, which enable all exclusive $VV$-pairs to be used except
   (probably) the fully hadronic ($VV\rightarrow JJJJ$) case. For example if $WW\rightarrow
   JJ\mu\nu$ then not only $M_{JJ} = M_W$ but \[M^2_{invisible} =
   (p_1+p_2-p_3-p_4-p_{J1}-p_{J2}-p_{\mu})^2 = M_\nu^2 = 0\] and 
   \[MM^2_2 = (p_1+p_2-p_3-p_4-p_{J1}-p_{J2})^2 = M_W^2.\] In the case $ZZ\rightarrow
   e^+e^-\nu\bar{\nu}$ then
   \[MM^2_{invisible} = (p_1+p_2-p_3-p_4-p_{e^-}-p_{e^+})^2 = M_Z^2\] which is an interesting
   case of seeing $Z\rightarrow \nu\bar{\nu}$ as a narrow peak (only using precision tracking,
   no calorimetry). Unfortunately it may be only a handful (or even no) events, but this
   makes it an almost background free channel for new (BSM) physics.
   
   The only new physics of which I am aware that should give a \emph{large peak} in the
   aforementioned $Z \rightarrow \nu \bar{\nu}$ mass plot is Alan White's theory of color sextet
   quarks and what I call the ``white pomeron". This would be a spectacular discovery. The white
   pomeron is a reggeized gluon color-neutralized by an infinite cloud of wee gluons. A pair of
   heavy color sextet (analogous to double color, like $RR$) quarks $U$ and $D$, are required
   to saturate asymptotic freedom. Double pomeron exchange will produce exclusive $W^+W^-$ and $ZZ$
   prolifically, once above threshold (at the LHC not the Tevatron) through $U,D$ loops. We can also have
   $Z$ photoproduction ($\gamma\pom \rightarrow Z$) at the LHC which would also be dramatic as it
   is tiny in the Standard Model. FP420~\cite{fp} has the potential for dramatic discoveries!
   
   \section{Summary}
   
   The Tevatron is, and the LHC will be, a cornucopia of QCD physics. At the LHC we will have
   the $Q^2$-range $\approx 0.01 \rightarrow 50,000,000$ GeV$^2$. We will measure jets, W, Z,
   b, c production, spectroscopy and decays, rapidity gaps, exclusive production
   of jets, $H,WW,ZZ$ and BSM physics. Adding very forward proton detection can open up a new
   window on strong and electroweak interactions.

\section*{Acknowledgments}
I thank the organizers of this excellent conference for inviting me, and Fermilab and the DOE
for support. Among colleagues in the $pXp$ business too numerous to mention, I thank especially
Valery Khoze and Alan White for theoretical inspiration.


\section*{References}
\begin{thebibliography}{99}
\bibitem{cdf} Space limitations prevent me from referring to the CDF and D0 papers
individually. The CDF publications can all be found at 
http://www-cdf.fnal.gov/physics/preprints/ 
\bibitem{dzero} The D0 publications can be found at
http://www-d0.fnal.gov/www-buffer/pub/publications.html 
\bibitem{fp} M.G.Albrow et al., FP420: An R\&D Proposal to Investigate the Feasibility of Installing
Proton Tagging Detectors in the 420m Region at LHC. CERN-LHCC-2005-025; LHCC-I-015.
\bibitem{kkmrs} V.A.Khoze et al., Diffractive Processes as a Tool for Searching for New Physics, 
hep-ph/0507040 and references therein.
\bibitem{albmm} M.G.Albrow and A.Rostovtsev, Searching for the Higgs Boson at Hadron Colliders using the Missing Mass
Method, hep-ph/0009336.
\bibitem{kmr1} V.A.Khoze,A.D.Martin,M.G.Ryskin, Eur.Phys.J C{\bf 19}, 477 (2001); erratum, ibid. C{\bf
20}, 599 (2001).
\bibitem{ell} J.R.Ellis, S.J.Lee and A.Pilaftsis, Phys.Rev. D{\bf 71} (2005) 075007.
\bibitem{arw} A.R.White, The Physics of a Sextet Quark Sector, hep-ph/0412062 and references therein.
\end{thebibliography}
\end{document}